# Electronic and magnetic transitions in perovskite SrRu$_{1-x}$Ir$_x$O$_3$ thin films

Abhijit Biswas, Yong Woo Lee and Yoon Hee Jeong[*]

Department of Physics, POSTECH, Pohang, 790-784, South Korea

**Abstract:**

We have investigated the electronic and magnetic properties of perovskite SrRu$_{1-x}$Ir$_x$O$_3$ ($0.0 \leq x \leq 0.25$) thin films grown by pulsed laser deposition on atomically-flat (001) SrTiO$_3$ substrates. SrRuO$_3$ has the properties of a ferromagnetic metal (resistivity $\rho \sim 200$ μΩ·cm at $T = 300$ K) with Curie temperature $T_C \sim 150$ K. Substituting Ir ($5d^{5+}$) for Ru ($4d^{4+}$) in SrRuO$_3$, films ($0.0 \leq x \leq 0.20$) showed fully-metallic behavior and ferromagnetic ordering, although $\rho$ increased and the ferromagnetic $T_C$ decreased. Films with $x = 0.25$ underwent the metal-to-insulator transition ($T_{MIT} \sim 75$ K) in $\rho$, and spin-glass-like ordering ($T_{SG} \sim 45$ K) with the elimination of ferromagnetic long-range ordering caused by the electron localization at the substitution sites. In ferromagnetic films ($0.0 \leq x \leq 0.20$), $\rho$ increased near-linearly with $T$ at $T > T_C$, but in paramagnetic film ($x = 0.25$) $\rho$ increased as $T^{3/2}$ at $T > T_{MIT}$. Moreover, observed spin-glass-like ($T_{SG}$) ordering with the negative magnetoresistance at $T < T_{MIT}$ in film with $x = 0.25$; validates the hypothesis that (Anderson) localization favors glassy ordering at amply disorder limit. These observations provide a promising approach for future applications and of fundamental interest in $4d$ and $5d$ mixed perovskites.

# I. INTRODUCTION

Tuning the electronic and magnetic ground states in strongly-correlated transition metal oxides (TMOs) have become important goals in condensed matter physics and materials science following the discoveries of metal-insulator transition (MIT), high-temperature superconductivity, and colossal magnetoresistance.[1-3] Among numerous transition metal oxides, $4d$ element $Ru^{4+}$-based transition-metal perovskite $SrRuO_3$ has many intriguing properties,[4] especially its low resistivity $\rho \sim 200$ μΩ·cm at room temperature, and ferromagnetism with Curie temperature $T_C \sim 150$ K and a saturation magnetic moment $1.1 \leq \mu_B/Ru^{4+} \leq 1.6$.[4-9] The origin of ferromagnetism in this compound remains subject to debate[10-12] because its saturation moment is less than the magnetic moment expected for the localized $t_{2g}^4$ low-spin configuration of $Ru^{4+}$ ions in an octahedral crystal field of O ions. Many authors believe that the ferromagnetism of perovskite $SrRuO_3$ is of itinerant (Stoner-type) nature because this structure predicts that in $SrRuO_3$, Coulomb interaction $U$ among itinerant electrons induces the spin split in the electronic band structure associated with the hybridization between the Ru ($t_{2g}$)–O ($2p$) orbitals which results in preferable spin directions.[13-18] $SrRuO_3$ can also be regarded as a bad metal.[7,19-21]

Transport and magnetic properties of perovskite $SrRuO_3$ can be tuned by chemical doping.[22-31] In particular, the Ru-site chemical doping effect in $SrRuO_3$ has been investigated by doping with a $3d$ magnetic ($Fe^{3+}$, $Mn^{3+}$, $Co^{2+}$, $Ni^{2+}$), a $3d$ non-magnetic ($Mg^{2+}$, $Sc^{3+}$, $Pb^{4+}$, $Ti^{4+}$), and a $4d$ element ($Rh^{4+}$) elements; results included suppression of the ferromagnetic $T_C$ and to the metal-insulator transition in the electronic transport. Doping with a $3d$ element hinders hybridization between Ru $4d$ and O $2p$ orbitals; in general this phenomenon localizes the charge carriers because $3d$ orbitals are generally more localized than are $4d$ orbitals. When Rh is used as a dopant, $4d$ $Rh^{4+}$ doping changes the electronic configuration from $4d^4$ to $4d^5$, and the Ru-site randomness has significant effects on the electronic and transport properties.[32]

Substituting a $5d$ based transition metal (let's say $Ir^{4+}$) element in the place of $Ru^{4+}$, in a $4d$ ($Ru^{4+}$) based ferromagnetic metal $SrRuO_3$ would be of fundamental interest in condensed matter physics, because transition metal oxides based on $5d$ elements show many correlated quantum phenomena due to interplay between electron correlation and spin-orbit coupling (SOC).[33] Consequently, a $5d$ element has a wider $d$ orbital, but weaker correlation ($U$) than do $3d$ or $4d$ elements, so replacing a $4d$ element with a $5d$ element should increase the band width $W$ and therefore decrease the overall effective interaction $U/W$. These changes might affect the overall magnetism, because $SrRuO_3$ has strong ferromagnetism of itinerant nature (band magnetism). Furthermore, the ionic radius of $Ru^{4+}$ is different from that of $Ir^{4+}$, so replacing $Ru^{4+}$ with $Ir^{4+}$ would change the *B-O-B* bond length and *B-O* bond angle (*B* is related to the $ABO_3$ perovskite); these changes would also affect $W$ and $U/W$. One fascinating characteristic of oxides based on heavy $5d$ elements is that the oxides have very strong SOC $\propto z^4$.[34] Therefore, substituting $Ir^{4+}$ for $Ru^{4+}$ will increase SOC, because Ir has $z = 77$ whereas Ru has $z = 44$. In this study, we assessed whether doping $5d$ element Ir into $SrRuO_3$ provides emergent phenomena not observed in systems doped with $3d$ or $4d$ elements. Since $SrRuO_3$ is nearly cubic perovskite and have wide-range of applications including as electrodes and junction layers in microelectronic devices,[35-37] it would also be desirable to have control over its functional properties from all possible aspects, considering the progress in superlattice thin film growth.

Here, we have systematically investigated the evolution of the electronic and magnetic properties in perovskite $SrRu_{1-x}Ir_xO_3$ ($0.0 \leq x \leq 0.25$) thin films. The characteristics of the perovskite were affected by $x$. Over the range $0.0 \leq x \leq 0.20$, $\rho$ increased and ferromagnetic $T_C$ decreased. At $x = 0.25$, the compound showed a low temperature metal-insulator transition in $\rho$, and ferromagnetic $T_C$ was fully suppressed with the observation of spin-glass like behavior at $T < T_{MIT}$. Interestingly, we find that although for all the ferromagnetically ordered films ($0.0 \leq x \leq 0.20$), above $T>T_C$, resistivity follows a near-linear $T$ dependence. In contrast, film with no long-range magnetic order ($x=0.25$) follows a $T^{3/2}$ temperature dependence in resistivity above $T>T_{MIT}$. Also negative MR was observed at $T < T_{MIT}$, supporting the localization picture with the coexistence of spin-glass ordering. Our effort would probably be a unique example to investigate the maneuver of the electronic and magnetic properties of perovskite SrRuO$_3$ thin films by doping a 5$d$ based transition metal element Ir on Ru-site.

## II. EXPERIMENTAL METHODS

We used pulsed laser deposition to grow high-quality epitaxial ~70-nm films of the SrRu$_{1-x}$Ir$_x$O$_3$ ($0.0 \leq x \leq 0.25$) on TiO$_2$-terminated SrTiO$_3$ (001) substrates. The system uses a KrF laser with a wavelength 248 nm, with laser energy density 2 J cm$^{-2}$ and a repetition rate of 4 Hz focused on the target. Substrate temperature was ~680 °C and O$_2$ partial pressure was 20 mTorr. The distance from the target to the substrate was ~50 mm. To compensate for oxygen vacancies, all films were cooled slowly (~7 °C/min) in the same O$_2$ partial pressure. For each compositional deposition, we used *different targets* made by stoichiometric mixing of SrCO$_3$, IrO$_2$ and RuO$_2$ raw powders and sintering them at $T = 1000$ °C at ambient pressure. Before growth, SrTiO$_3$ substrates (miscut < 0.1°), were made to be atomically flat with one unit-cell step height by using conventional etching and annealing process.[38] Each film's epitaxial pseudo-cubic phase was characterized by an x-ray diffraction (XRD) $\theta$-$2\theta$ scan using a Diffractometer (Rigaku). The temperature-dependent $\rho$ of each film was measured using the four-terminal van der Pauw transport measurement technique performed with a Physical Property Measurement System (Quantum Design) at temperatures 10 K ≤ $T$ ≤ 300 K. Magnetization of each film was measured using a Physical Property Measurement System in field-cooled (FC) and zero-field-cooled (ZFC) configurations at applied field strength of 1000 Oe. Magnetoresistance was measured with magnetic field magnitude up to ± 9 T, applied perpendicular to the film surface.

## III. RESULTS AND DISCUSSION

Structurally, SrRuO$_3$ crystallizes in an orthorhombic distorted GdFeO$_3$-type perovskite structure (space group *Pbnm*) with lattice parameters $a$ = 5.53 Å, $b$ = 5.57 Å, and $c$ = 7.85 Å at room temperature.[39] In contrast, the analogous perovskite structure of SrIrO$_3$ (space group *Pbnm*) with lattice parameters $a$ = 5.56 Å, $b$ = 5.58 Å, and $c$ = 7.89 Å can only be formed at high temperature and pressure.[40] The pseudocubic ($a_c$) lattice parameter of bulk perovskite SrRuO$_3$ converted from the orthorhombic lattice parameters is ~3.93 Å and the perovskite SrIrO$_3$ pseudo-cubic ($a_c$) lattice constant is ~3.96 Å. XRD *θ-2θ* scans of the SrRu$_{1-x}$Ir$_x$O$_3$ thin films (0.0 ≤ $x$ ≤ 0.25) on the SrTiO$_3$ (001) substrates (Fig. 1a) show only film and substrate peaks of perovskite origin, and no impurity peaks. As $x$ increased, the *2θ* value of the film peak maximum decreased (Fig. 1b); i.e., the film's out-of-plane lattice constant increased assuming its in-plane lattice constants are locked. This change is plausible as according to Vegard's law, at constant temperature, the crystallographic parameters of a continuous substitutional compound vary linearly with the concentration.[41] Considering that all films were grown at the *same* temperature and O$_2$ partial pressure, these results indicate that as $x$ increased, pseudo-cubic lattice constants and the overall unit cell volumes also increased. The films also show uniform layer thickness fringes (Fig. 1b), which confirm good crystalline quality, consistent thickness of the film, and homogeneous nature of the film surface. Ambient atomic force microscopy (AFM) shows (not shown here) that films surfaces were nearly flat with roughness ∼ 1-2 nm.

Having confirmed the good crystalline quality, we then measured $\rho$ of all samples (Fig. 2). The parent SrRuO$_3$ had $\rho \sim 210$ μΩ·cm at $T = 300$ K and a ferromagnetic transition at $T_C \sim 150$ K, which is signaled by the presence of an anomaly in $\rho$ at $T \sim 150$ K, which we attribute to the ferromagnetic ordering. These numbers are close to the reported value of $\rho \sim 190$ μΩ·cm at $T = 300$ K for bulk high-quality single crystals of SrRuO$_3$ **[6]**. Irrespective of the origin of the ferromagnetism, doping with Ir decreased $T_C$ from ∼150 K ($x$=0.00) to ∼90 K ($x$=0.20) (Fig. 2, arrows). Doping with Ir increased $\rho$ also, but films with $x \leq 0.20$ showed fully metallic behavior (decrease of $\rho$ with decrease in temperature) down to $T = 10$ K; this trend indicates that our films do not suffer severely from disorder due to flaws as such as grain boundaries and step edges, which may cause even parent SrRuO$_3$ to show an increase in $\rho$ at low temperature **[42]**, although there always be a possibility of crystallographic disorder due to atomic substitution, since increasing Ir concentration increases the resistivity.

At $x = 0.25$, $\rho$ increased to ∼1000 μΩ·cm at $T = 300$ K, and low-temperature metal-insulator transition in $\rho$ occurred at $T_{\text{MIT}} \sim 75$ K (Fig. 2). A similar feature has been observed in Ir-doped SrRuO$_3$ films with $x = 0.50$ ($T_{\text{MIT}} \sim 80$ K).**[43]** More importantly, the film with $x = 0.25$ did not show kink or unusual behavior in $\rho$ over the temperature range 10 K $\leq T \leq$ 300 K; i.e., presumably the film did not have any long-range ferromagnetic ordering. This observation is substantiated by the temperature derivative of resistivity (TDR) curves $d\rho/dT$ (Fig. 3): TDRs of all SrRu$_{1-x}$Ir$_x$O$_3$ thin films with $x \leq 0.20$ showed an anomaly (Fig. 3a) consistent with long-range magnetic ferromagnetic ordering, but no anomaly in TDR was observed in the film with $x = 0.25$ (Fig. 3b); this lack confirms that the film has no long-range magnetic order. Also quite interesting to notice that in the film with $x$=0.25, the sign of the TDR changed at $T \sim 75$ K; this change has been associated empirically with a phase transition from metal-to-insulator.[44]

The variation in $\rho$ vs. $T$ in SrRu$_{1-x}$Ir$_x$O$_3$ thin films is interesting, because $\rho$ of SrRuO$_3$ changes almost linearly with $T$ even at higher temperature, and passes through the Mott–Ioffe–Regel limit.[21] This almost-linear $T$ dependence has been attributed to electron-phonon scattering at $T > T_C$. Also, SrRuO$_3$ is apparently a Fermi liquid at very low $T < T_C$; this transition has been attributed to the scattering mechanism of weakly interacting particles.[19] Therefore, considering the fact of ferromagnetic to paramagnetic crossover in mind, we analyzed the temperature-dependence of $\rho$ in the one film with long-range ferromagnetic ordering ($x$=0.10) and one with no long-range magnetic ordering ($x$=0.25).

In the film with long-range ferromagnetic ordering ($x$=0.10) (Fig. 4a) the relationship between $\rho$ and $T$ was almost linear at $T > T_C$, and showed Fermi liquid behavior at 10 K $\leq T \leq$ 30 K. Fitting the temperature dependent resistivity to the typical formula $\rho = \rho_0 + AT^2$ (inset, Fig. 4a) yielded residual resistivity $\rho_0 = 0.30986$ m$\Omega \cdot$cm which measures scattering due to disorder, and temperature coefficient $A = 1.426 \times 10^{-8}$ $\mu\Omega \cdot$cm$\cdot$K$^{-2}$ which represents the scattering strength. Variations in $\rho$ were similar in the other ferromagnetically-ordered films ($x \leq 0.20$; data not shown). These results in SrRu$_{1-x}$Ir$_x$O$_3$ thin films with $x \leq 0.20$ are consistent with the reported behavior at $T \geq T_C$ (near-linear $T$ dependence) and at $T \leq T_C$ (Fermi liquid) for ferromagnetic SrRuO$_3$.[19]

In contrast, in the film with no long-range magnetic ($x = 0.25$), $\rho$ follows a $T^{3/2}$ dependence at 300 K $\geq T \geq T_{MIT}$. Resistivity fitting yielded $A \sim 5.11 \times 10^{-8}$ $\Omega \cdot$cm$\cdot$K$^{-1.5}$, which is close to the value of $\sim 3 - 5 \times 10^{-8}$ $\Omega \cdot$cm$\cdot$K$^{-1.5}$ that is common in other strongly-correlated electron systems such as Na$_{0.5}$CoO$_2$ and CaVO$_3$.[45] The increase in $\rho$ at $T \leq 25$ K can be fitted with the three-dimensional variable range hopping model (i.e., $ln(\sigma) \propto 1/T^{1/4}$; inset, Fig. 4b). Increase in $x$ in SrRu$_{1-x}$Ir$_x$O$_3$ thin films from $x = 0.20$ to $x = 0.25$ eliminated the long-range ferromagnetic order, and changed the relationship between $\rho$ and $T$ as it changed from $\rho \propto T$ to $\rho \propto T^{3/2}$. This $\rho \propto T^{3/2}$ relationship in the metallic region (300 K $\geq T \geq T_{MIT}$) and three-dimensional variable range hopping fitting at $T \leq T_{MIT}$ have also been observed in our recent report in perovskite SrRu$_{0.5}$Ir$_{0.5}$O$_3$ and SrIrO$_3$ thin films grown on (001) SrTiO$_3$ substrates.**[43,46]** This $T^{3/2}$ dependence of $\rho$ at $x = 0.25$ may occur because as the system goes through the transition to the Mott-Hubbard regime from the itinerant side, the quasi-particle scenario breaks down and non-Fermi liquid resistivity ($\rho \propto T^{3/2}$) is predicted to be associated with the scattering between localized electrons by locally cooperative bond-length fluctuations and itinerant electrons.**[45]** We observed no spin fluctuation (ferro or anti-ferro) (next section), and $T_{MIT}$ was quite high ($T_{MIT} \sim 75$ K) in our system; for these reasons we can exclude the possibilities of (1) self-consistent renormalization based on *spin fluctuations* mechanism for the $T^{3/2}$ variation in resistivity, (2) the contribution of the incoherent part of electron magnon scattering to resistivity showing a $T^{3/2}$ dependent at *low temperature*, and (3) the *low temperature* Kondo effect for upturn in resistivity.**[47-49]**

To confirm the systematic variation in magnetic ordering in our samples as observed in the resistivity, we measured the magnetization as a function of temperature for $0.0 \leq x \leq 0.25$ thin films at applied magnetic field along perpendicular to the plane (i.e., out-of-plane). In the field-cooled (FC) measurements, magnetic ordering temperature decreased as $T_C$ changed from ~150 K at $x = 0.00$ to ~90 K at $x = 0.20$ (Fig. 5). Doping with Ir magnetic impurity seems to have caused increase in magnetic disorder due to the randomness between Ir$^{4+}$ and Ru$^{4+}$ moments, which hinders the hybridization between Ru 4$d$ orbitals and O 2$p$ orbitals and localizes the carrier. This localization depletes the DOS near $E_F$ and thereby suppresses the Stoner instability and the reduction in the ferromagnetic $T_C$.[31]

Increasing $x$ to 0.25 did not cause any long-range ferromagnetic ordering, because it was fully suppressed (more magnetic impurity related disorder effect). This result is consistent with the lack of a kink in the resistivity curve (Fig. 2). But even in this scenario, the possibilities of local magnetic ordering and freezing of ferromagnetic clusters cannot be ignored. The magnetization measurement (Fig. 6a) for $x = 0.25$ film shows a visible cusp in the zero-field-cooled (ZFC) configuration at $T_{SG} \sim 45$ K; this is characteristic of spin-glass-like behavior. The possibility that this spin-glass effect is caused by the substrate can be excluded, because the SrTiO$_3$ substrate is diamagnetic. Similar spin-glass-like behavior at $T_{SG}$ ~ 25 K has also been observed in SrRu$_{0.5}$Ir$_{0.5}$O$_3$ thin film.[43] The very weak cusp with almost flat magnetization curve can be attributed with: (1) the moment of the film was very low (as expected for a system containing a 5$d$ element that has weak correlation) even at low temperature (with respect to parent SrRuO$_3$), (2) although there is no mean ferromagnetism as the interaction of spins dissolved in a matrix, but there might exist a ground state with spins aligned in a particular direction and hence the applied magnetic field is not adequate to align all the domains along the direction of the field.[50] It should stressed that spin-glass is a nontrivial phenomena which arises due to competitive interactions between disorders and magnetic frustration and cannot be described by a single parameter. Below this spin-glass ordering temperature, at $T = 10$ K, the film had negative magnetoresistance (Fig. 6b). In general, shrinkage of the localized electron wave function results in a decrease in the hopping rate and an increase in $\rho$, and together these changes result in a positive MR.[51] This wave function shrinkage mechanism cannot explain the observed negative MR in our system. Therefore the observed negative MR at $T \leq T_{MIT}$ may be related to the weak localization quantum interference mechanism.[52] This coexistence of the spin-glass ordering and the insulating region in $\rho$ is consistent with the hypothesis that the (Anderson) localization favors glassy ordering at amply disorder limit.[53]

At present, a complete microscopic description of the nature of above transitions is lacking, but the Ir-doping effect in SrRuO$_3$, origin of disorder and thus the suppression of the ferromagnetic $T_C$ can be understood qualitatively within the context of Anderson-Hubbard Hamiltonian picture, because it considers all relevant energy terms including $U$, hopping integral between neighboring sites, and a random site potential. This model can explain the destruction of coherent scattering as well as the breakdown of the quasi-particle scenario which decreases the density of states near level $E_F$ and hence decreases the ferromagnetic $T_C$. [31] Doping the SrRuO$_3$ with a 3$d$ or 4$d$ element localizes the carriers and induces disorder (which results in the metal-insulator transition and suppression of $T_C$) would be somewhat similar considering its almost similar energy scale of $U$, $W$ and SOC. In general, SrRuO$_3$ has no local magnetic moment, and, therefore, may be considered as a classical itinerant ferromagnet. Based on above facts, replacing Ir$^{4+}$ in Ru$^{4+}$ site, it is then difficult to understand that magnetism in Ir doped SrRuO$_3$ changes to local ferromagnet (or spin-glass-like) because Ir$^{4+}$ (5$d$) is more spatially extended than Ru$^{4+}$ (4$d$). However, a 5$d$ element might induce an extra term in the Hamiltonian because a 5$d$ element has large spin-orbit coupling. We presume that strong spin-orbit coupling (w. r. t. 3$d$ and 4$d$) and low on-site Coulomb interaction (w. r. t. 3$d$ and 4$d$) of 5$d$ element compromise the relevant effect in the band picture. If this hypothesis is correct, the 5$d$ doping effect in SrRuO$_3$ should be similar to the doping effect of 3$d$ or 4$d$ systems. This possibility can be tested by spectroscopy and optical conductivity measurements. Apart from magnetic disorder, we cannot neglect other possibilities of disorder due to the structural defects, vacancies, stoichiometry variations, and the inter-diffusion between the film-substrate interfaces which originates due to the release of strain energy in the highly strained thin film, and these are obvious facts which should be kept in mind for further experimentations.

## IV. CONCLUSIONS

We grew high-quality epitaxial thin films of perovskite SrRu$_{1-x}$Ir$_x$O$_3$ ($0.0 \leq x \leq 0.25$) on atomically-flat SrTiO$_3$ (001) substrates, and measured the films' electronic and magnetic properties. Resistivity measurements showed fully metallic behavior for films with $0.0 \leq x \leq 0.20$, but a low temperature metal-insulator transition ($T_{MIT} \sim 75$ K) in film at $x = 0.25$ due to the disorder-related carrier localization effect. Simultaneously, the ferromagnetic $T_C$ decreased as $x$ increased and the ferromagnetic long-range ordering disappeared at $x = 0.25$. At $x = 0.25$, the SrRu$_{1-x}$Ir$_x$O$_3$ showed spin-glass like behavior at $T_{SG} \sim 45$ K which is below the metal-insulator transition temperature ($T_{MIT} \sim 75$ K). Resistivity in the paramagnetic region ($T \geq T_C$) followed a near-linear $T$ dependence in all ferromagnetically ordered films ($0.0 \leq x \leq 0.20$), but a $T^{3/2}$ dependence at $x = 0.25$ that has no long-range magnetic order. These observations suggest that disorder induced by the random distribution of Ir$^{4+}$ magnetic impurities over Ru$^{4+}$ sites hinders the hybridization between the Ru (t$_{2g}$)–O (2$p$) orbitals and localizes the carriers, thus reducing the ferromagnetic $T_C$. The film with $x = 0.25$ had a low metal-insulator transition ($T_{MIT} \sim 75$ K) temperature and glassy-like ordering ($T_{SG} \sim 45$ K) with negative magnetoresistance at $T=10$ K; these characteristics are consistent with the hypothesis that the (Anderson) localization favors the glassy phase in a strongly-correlated system at the certain disorder limit.

Above results provide basic information on the effects of replacing 4$d$ Ru$^{4+}$ with 5$d$ Ir$^{4+}$ in perovkite SrRuO$_3$. Further microscopic analysis of the films, especially near the ferromagnetic to spin–glass crossover region might provide insight into the mechanisms that cause the observed behaviors of the films. Spectroscopic measurements might help to quantify the effects of each element and the magnetic behavior. The dependence of $T_C$ on $x$ observed in our study should be sought in other substrates also to determine whether it is universal or instead depends on the amount of the strain (e.g., $T_C$ vs. $x$ vs. *strain*) that is caused by the film-substrate lattice mismatch, because strain always has a strong influence on thin film functionalities. Although, our macroscopic transport measurements provide only qualitative insights into the origin and nature of the evolution of the electronic and magnetic state, but they can guide further investigations to understand the physical origin of the electronic and magnetic properties caused by $5d$ transition metal doping of strongly correlated oxides, and may enable novel device applications and new emergent phenomena, especially from the $SrRuO_3$-$SrIrO_3$ superlattice perspectives.

## ACKNOWLEDGMENTS

We thank Prof. B. I. Min and Prof. J. S. Kang for useful discussions. This work was supported by the National Research Foundation via SRC at POSTECH (2011-0030786).

## REFERENCES

[1] M. Imada, A. Fujimori, and Y. Tokura, Rev. Mod. Phys. **70**, 1039 (1998). (http://dx.doi.org/10.1103/RevModPhys.70.1039)

[2] J. B. Goodenough, *Localized to Itinerant Electronic Transition in Perovskite Oxides* (Springer Series in Structure and Bonding, Vol. 98, Berlin: Springer, 2001)

[3] J. B. Goodenough and J.-S. Zhou, Chem. Mater. **10**, 2980 (1998). (10.1021/cm980276u)

[4] G. Koster, L. Klein, W. Siemons, G. Rijnders, J. S. Dodge, C.-B. Eom, D. H. A. Blank, and M. R. Beasley, Rev. Mod. Phys. **84**, 253 (2012) and references therein (10.1103/RevModPhys.84.253)

[5] J. M. Longo, P. M. Raccah, and J. B. Goodenough, J. Appl. Phys. **39**, 1327 (1968). (10.1063/1.1656282)

[6] G. Cao, S. McCall, M. Shepard, J. E. Crow, and R. P. Guertin, Phys. Rev. B **56**, 321 (1997). (10.1103/PhysRevB.56.321)

[7] P. B. Allen, H. Berger, O. Chauvet, L. Forro, T. Jarlborg, A. Junod, B. Revaz, and G. Santi, Phys. Rev. B **53**, 4393 (1996). (10.1103/PhysRevB.53.4393)

[8] S. N. Bushmeleva, V. Yu. Pomjakushin, E. V. Pomjakushina, D. V. Sheptyakov, and A. M. Balagurov, J. Magn. Magn. Mater. **305**, 491 (2006). (doi:10.1016/j.jmmm.2006.02.089)

[9] P. S. Anil Kumar, P. A. Joy, and S. K. Date, Physica B **269**, 356 (1999). (doi:10.1016/S0921-4526(99)00126-X)

[10] D. E. Shai, C. Adamo, D. W. Shen, C. M. Brooks, J. W. Harter, E. J. Monkman, B. Burganov, D. G. Schlom, and K. M. Shen, Phys. Rev. Lett. **110**, 087004 (2013). (10.1103/PhysRevLett.110.087004)

[11] D. W. Jeong, H. C. Choi, C. H. Kim, S. H. Chang, C. H. Sohn, H. J. Park, T. D. Kang, D.-Y. Cho, S. H. Baek, C. B. Eom, J. H. Shim, J. Yu, K. W. Kim, S. J. Moon, and T. W. Noh, Phys. Rev. Lett. **110**, 247202 (2013). (10.1103/PhysRevLett.110.247202)

[12] M. Kim, B. I. Min, arXiv preprint arXiv:1502.06322.

[13] E. C. Stoner, Proc. Phys. Soc. A **165**, 372 (1938). (10.1098/rspa.1938.0066)

[14] D. J. Singh, J. Appl. Phys. **79**, 4818 (1996). (10.1063/1.361618)

[15] I. I. Mazin and D. J. Singh, Phys. Rev. B **56**, 2556 (1997). (http://dx.doi.org/10.1103/PhysRevB.56.2556)

[16] K. Fujioka, J. Okamoto, T. Mizokawa, A. Fujimori, I. Hase, M. Abbate, H. J. Lin, C. T. Chen, Y. Takeda, and M. Takano, Phys. Rev. B **56**, 6380 (1997). (10.1103/PhysRevB.56.6380)

[17] N. Hiraoka, M. Itou, A. Deb, Y. Sakurai, Y. Kakutani, A. Koizumi, N. Sakai, S. Uzuhara, S. Miyaki, H. Koizumi, K. Makoshi, N. Kikugawa, and Y. Maeno, Phys. Rev. B **70**, 054420 (2004). (10.1103/PhysRevB.70.054420)

[18] K. Maiti, Phys. Rev. B **73**, 235110 (2006). (10.1103/PhysRevB.73.235110)

[19] L. Klein, J. S. Dodge, C. H. Ahn, J. W. Reiner, L. Mieville, T. H. Geballe, M. R. Beasley, and A. Kapitulnik, J. Phys.: Condens. Matter **8**, 10111 (1996). (doi:10.1088/0953-8984/8/48/026)

[20] N. E. Hussey, K. Takenaka, and H. Takagi, Philos. Mag. **84**, 2847 (2004). (10.1080/14786430410001716944)

[21] O. Gunnarsson, M. Calandra, and J. E. Han, Rev. Mod. Phys. **75**, 1085 (2003). (http://dx.doi.org/10.1103/RevModPhys.75.1085)

[22] C. Bansal, H. Kawanaka, R. Takahashi, and Y. Nishihara, J. Alloys Cpds. **360**, 47 (2003). (http://dx.doi.org/10.1016/S0925-8388(03)00333-5)

[23] J. Fan, S. Liao, W. Wang, L. Zhang, W. Tong, L. Ling, B. Hong, Y. Shi, Y. Zhu, D. Hu, L. Pi and Y. Zhang, J. Appl. Phys. **110**, 043907 (2011). (http://dx.doi.org/10.1063/1.3624764)

[24] A. J. Williams, A. Gillies, J. P. Attfield, G. Heymann, H. Huppertz, M. J. Martinez-Lope, and J. A. Alonso, Phys. Rev. B **73**, 104409 (2006). (10.1103/PhysRevB.73.104409)

[25] X.-Y. Zhang, Y. Chen, H.-X. Cao, Z.-Y. Li, and C. K. Ong, Solid State Commun. **145**, 259 (2008). (doi:10.1016/j.ssc.2007.11.017)

[26] L.Pi, A. Maignan, R. Retoux and B. Raveau, J. Phys.: Condens. Matter **14**, 7391 (2002). (doi:10.1088/0953-8984/14/31/310)

[27] I. Qasim, B. J Kennedy, Z. Zhang, M. Avdeev and L. −Y. Jang, J. Phys.: Condens. Matter 23, 435401 (2011). (doi:10.1088/0953-8984/23/43/435401)

[28] D. A. Crandles, M. M. Yazdanian, and F. S. Razavi, J. Phys. D: Appl. Phys. **39, 6** (2006). (doi:10.1088/0022-3727/39/1/002)

[29] W. Tong, F.-Q. Huang, and I.-W. Chen, J. Phys.: Condens. Matter **23**, 086005 (2011). (10.1088/0953-8984/23/8/086005)

[30] G. Cao, S. McCall, J. Bolivar, M. Shepard, F. Freibert, P. Henning, J. E. Crow, and T. Yuen, Phys. Rev. B **54**, 15144 (1996). (10.1103/PhysRevB.54.15144)

[31] K. W. Kim, J. S. Lee, T. W. Noh, S. R. Lee, and K. Char, Phys. Rev. B **71**, 125104 (2005). (10.1103/PhysRevB.71.125104)

[32] K. Yamaura, D. P. Young, and E. Takayama-Muromachi, Phys. Rev. B **69**, 024410 (2004). (http://dx.doi.org/10.1103/PhysRevB.69.024410)

[33] W. Witczak-Krempa, G. Chen, Y.-B. Kim, and L. Balents, Condens. Matter Phys. **5**, 2014 (2013). (10.1146/annurev-conmatphys-020911-125138)

[34] L. F. Mattheiss, Phys. Rev. B **13**, 2433 (1976). (10.1103/PhysRevB.13.2433)

[35] F. Pulizzi, Nat. Mater. **11**, 564 (2012). (doi:10.1038/nmat3375)

[36] M. Bibes and A. Barthelemy, IEEE Trans. Electron Devices **54**, 1003 (2007). (10.1109/TED.2007.894366)

[37] G. Herranz, B. Martinez, J. Fontcuberta, F. Sanchez, M. V. Garcia-Cuenca, C. Ferrater, and M. Varela, J. Appl. Phys. **93**, 8035 (2003). (http://dx.doi.org/10.1063/1.1555372)

[38] A. Biswas, P. B. Rossen, C.-H. Yang, W. Siemons, M.-H. Jung, I. K. Yang, R. Ramesh, and Y. H. Jeong, Appl. Phys. Lett. **98**, 051904 (2011). (http://dx.doi.org/10.1063/1.3549860)

[39] C. B. Eom, R. J. Cava, R. M. Fleming, Julia M. Phillips, R. B. vanDover, J. H. Marshall, J. W. P. Hsu, J. J. Krajewski, and W. F. Peck Jr., Science, **258**, 1766 (1992). (10.1126/science.258.5089.1766)

[40] J. M. Longo, J. A. Kafalas, and R. J. Arnott, J. Solid State Chem. **3**, 174 (1971). (doi:10.1016/0022-4596(71)90022-3)

[41] A. R. Denton and N. W. Ashcroft, Phys. Rev. A 43, 3161 (1991). (http://dx.doi.org/10.1103/PhysRevA.43.3161)

[42] G. Herranz, B. Martinez, J. Fontcuberta, F. Sanchez, C. Ferrater, M. V. Garcia-Cuenca, and M. Varela, Phys. Rev. B **67**, 174423 (2003). (http://dx.doi.org/10.1103/PhysRevB.67.174423)

[43] A. Biswas, Y. W. Lee, S. W. Kim, and Y. H. Jeong, J. Appl. Phys. **117**, 115304 (2015). (http://dx.doi.org/10.1063/1.4915943)

[44] S. V. Kravchenko and T. M. Klapwijk, Phys. Rev. Lett. **84**, 2909 (2000). (http://dx.doi.org/10.1103/PhysRevLett.84.2909)

[45] F. Rivadulla, J. S. Zhou, and J. B. Goodenough, Phys. Rev. B **67**, 165110 (2003). (10.1103/PhysRevB.67.165110)

[46] A. Biswas, K. -S. Kim, and Y. H. Jeong, J. Appl. Phys. **116**, 213704 (2014). (http://dx.doi.org/10.1063/1.4903314)

[47] D. L. Mills, A. Fert, and I. A. Campbell, Phys. Rev. B **4**, 196 (1971). (http://dx.doi.org/10.1103/PhysRevB.4.196)

[48] T. Moriya, *Spin Fluctuations in Itinerant Electron Magnetism* (Springer Series in Solid State Sciences 56, Berlin: Springer, 1985).

[49] L. Kouwenhoven and L. Glazman, Phys. World **14**, No. 1, 33–38 (2001).

[50] R. Palai, H. Huhtinen, J. F. Scott, and R. S. Katiyer. Phys. Rev. B **79**, 104413 (2009). (http://dx.doi.org/10.1103/PhysRevB.79.104413)

[51] O. Faran and Z. Ovadyahu, Phys. Rev. B **38**, 5457 (1988). (10.1103/PhysRevB.38.5457)

[52] A. Kawabata, Solid State Commun. **34**, 431 (1980). (10.1016/0038-1098(80)90644-4)

[53] V. Dobrosavljević, D. Tanaskovic, and A. A. Pastor, Phys. Rev. Lett. **90**, 016402 (2003). (10.1103/PhysRevLett.90.016402)

**FIGURE CAPTIONS**

**FIG. 1.** (Color Online) (a) X-ray diffraction $\theta$-$2\theta$ scans of perovskite SrRu$_{1-x}$Ir$_x$O$_3$ ($0.0 \leq x \leq 0.25$) thin films grown on atomically-flat (001) SrTiO$_3$ substrates. (b) As the Ir concentration ($x$) increases, $(001)_C$ peak position shifts toward the lower angle values (dashed vertical line). This means that with the increase $x$, pseudo-cubic lattice constants and overall unit cell volumes increase assuming a linear relationship between the end members $a_c \sim 3.93$ Å (perovskite SrRuO$_3$) and $a_c \sim 3.96$ Å (perovskite SrIrO$_3$).

**FIG. 2.** (Color Online) Resistivity of SrRu$_{1-x}$Ir$_x$O$_3$ thin films ($0.0 \leq x \leq 0.25$). At $x \leq 0.20$, films show fully metallic behavior within 10 K $\leq T \leq$ 300 K. At $x = 0.25$, a metal-to-insulator transition occurred at $T_{\text{MIT}} \sim 75$ K due to the localization effect (magnetic impurity disorder effect). Pink arrows: ferromagnetic transition for $0.0 \leq x \leq 0.20$ films; ferromagnetic $T_C$ decreases as $x$ increases, and does not occur at $x = 0.25$.

**FIG. 3.** (Color Online) (a) Temperature derivative of resistivity (TDR) shows anomalies at the ferromagnetic transitions temperature in SrRu$_{1-x}$Ir$_x$O$_3$ ($0.0 \leq x \leq 0.20$) thin films. (b) TDR curve of film at $x$=0.25 did not show any anomaly related to ferromagnetic transition.

**FIG. 4.** (Color Online). Resistivity $\rho$ vs. temperature $T$ of SrRu$_{1-x}$Ir$_x$O$_3$ films (a) $x = 0.10$: the film was ferromagnetic with $\rho \propto T$ at $T < T_C$, and showed Fermi liquid behavior at $T < T_C$ and within 10 K $\leq T \leq$ 30 K (inset). (b) For $x = 0.25$: the film had $\rho \propto T^{3/2}$ at $T \geq T_{\text{MIT}}$, and showed three-dimensional variable range hopping behavior ($\ln \sigma \propto 1/T^{1/4}$) within 10 K $\leq T \leq$

25 K.

**FIG. 5.** (Color Online) Field-cooled (FC) magnetization measurements of SrRu$_{1-x}$Ir$_x$O$_3$ ($0.0 \leq x \leq 0.20$) thin films. With increasing $x$, ferromagnetic $T_C$ decreases as the magnetic disorder due to Ir doping on Ru site becomes increasingly random and increasingly hinders hybridization between Ru (t$_{2g}$)–O (2$p$) orbitals, and thereby localizes the carriers.

**FIG. 6.** (Color Online) (a) Zero-field-cooled (ZFC) and field-cooled (FC) magnetization measurements of the SrRu$_{0.25}$Ir$_{0.75}$O$_3$ film. Visible cusp at $T_{SG} \sim 45$ K in ZFC measurement is the signature of spin-glass-like ordering. (b) SrRu$_{0.25}$Ir$_{0.75}$O$_3$ film had transverse negative magnetoresistance at $T = 10$ K at magnetic field $-9 \leq B \leq 9$ T.

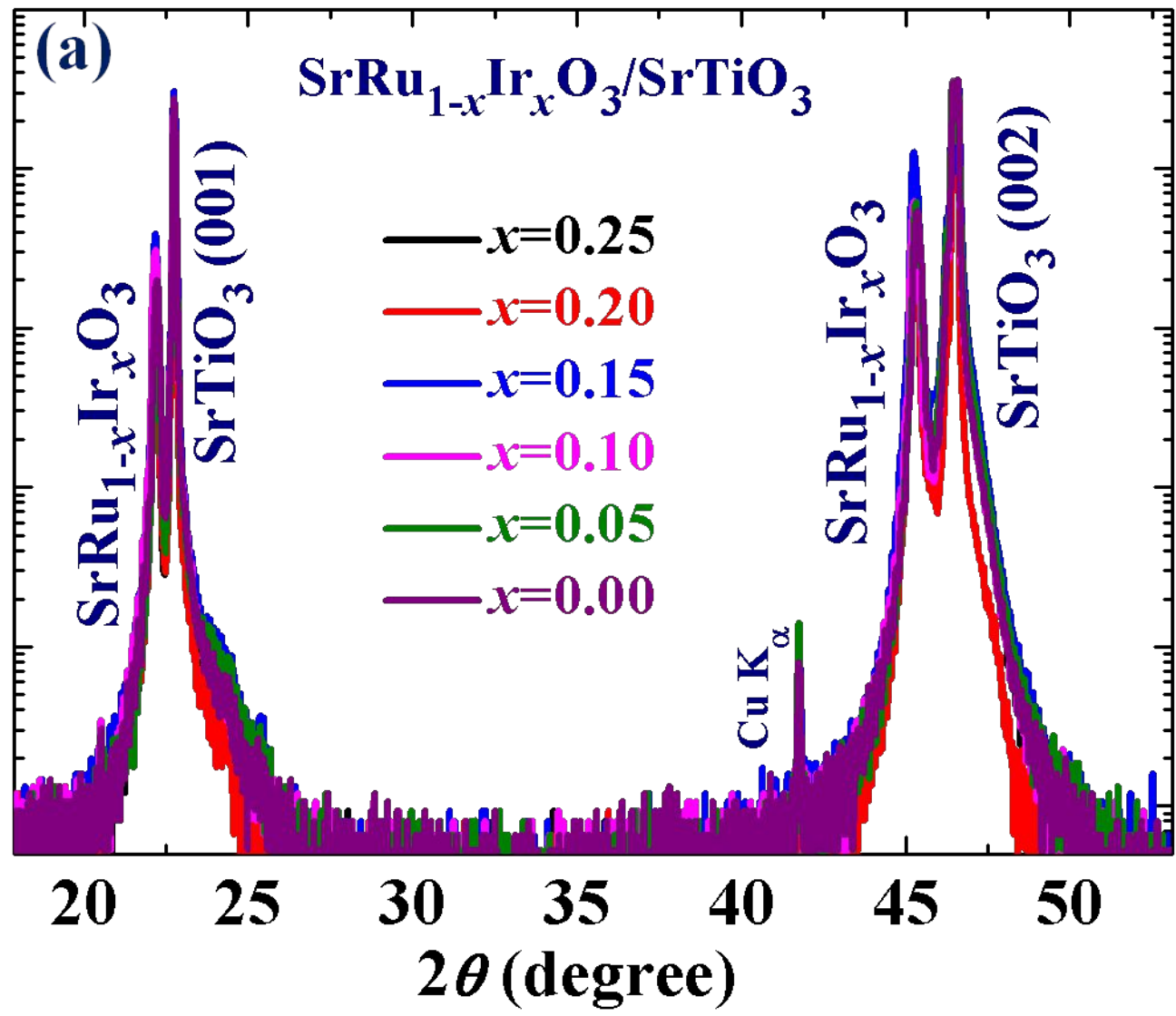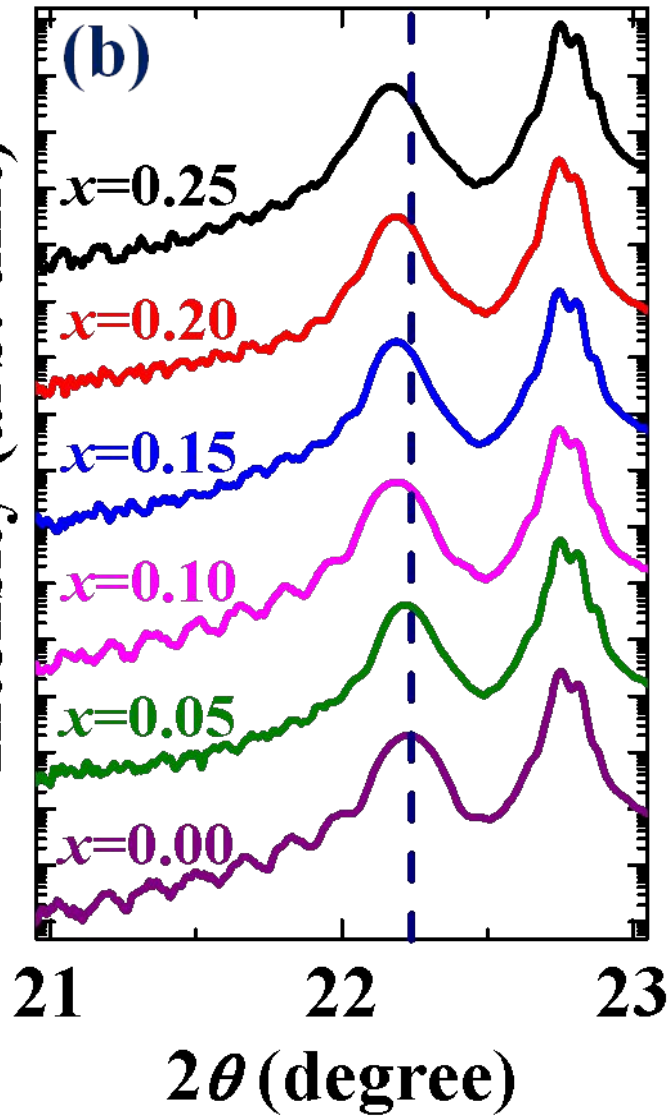

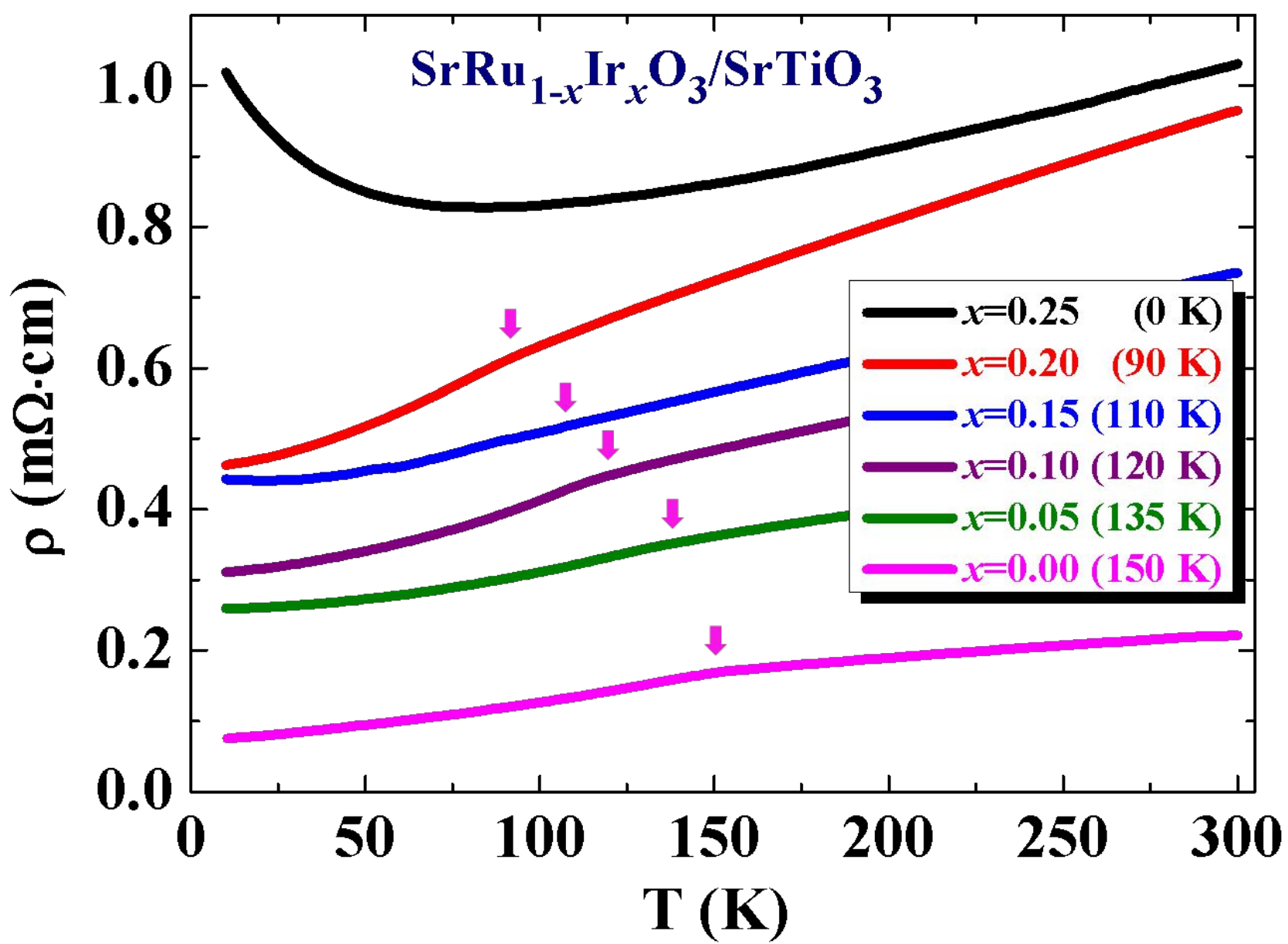

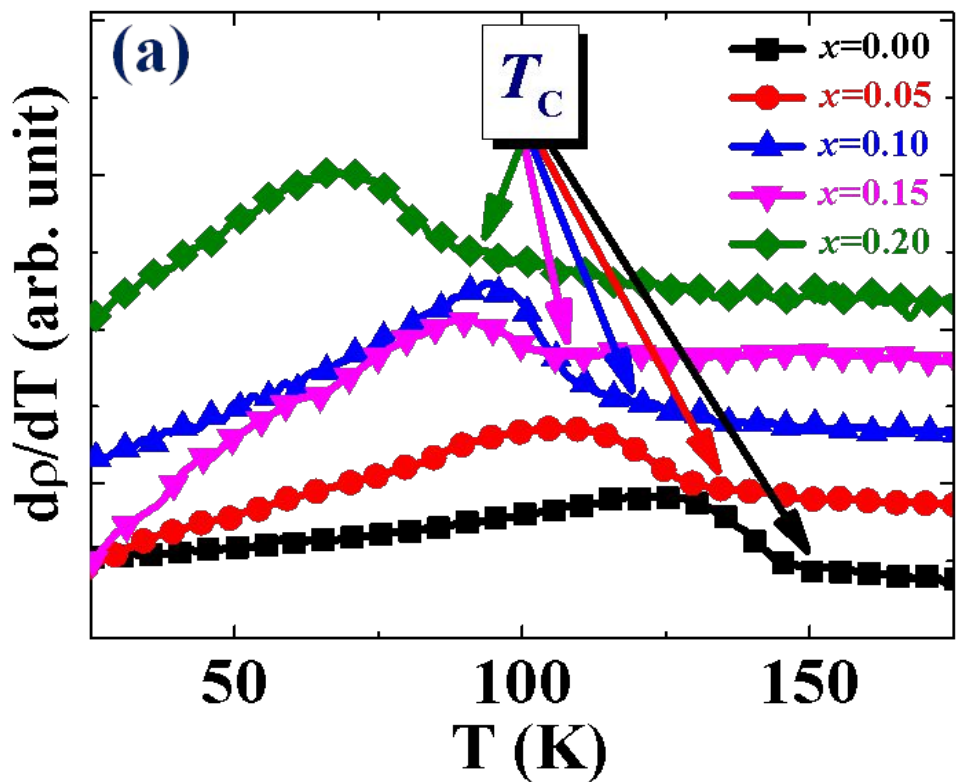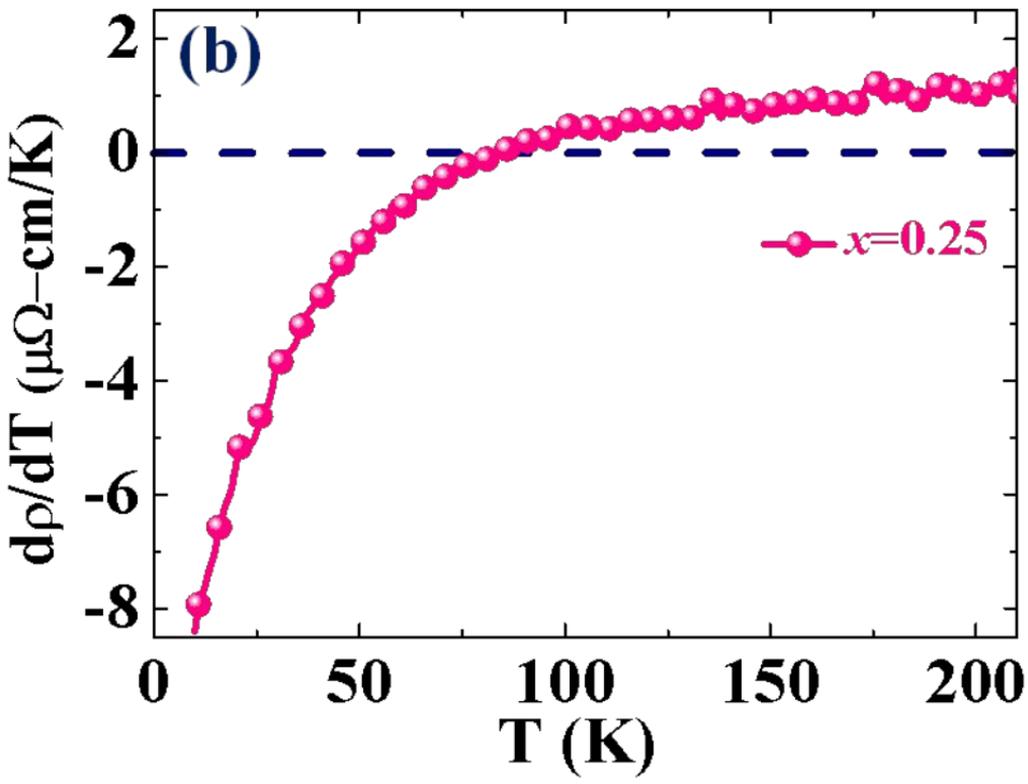

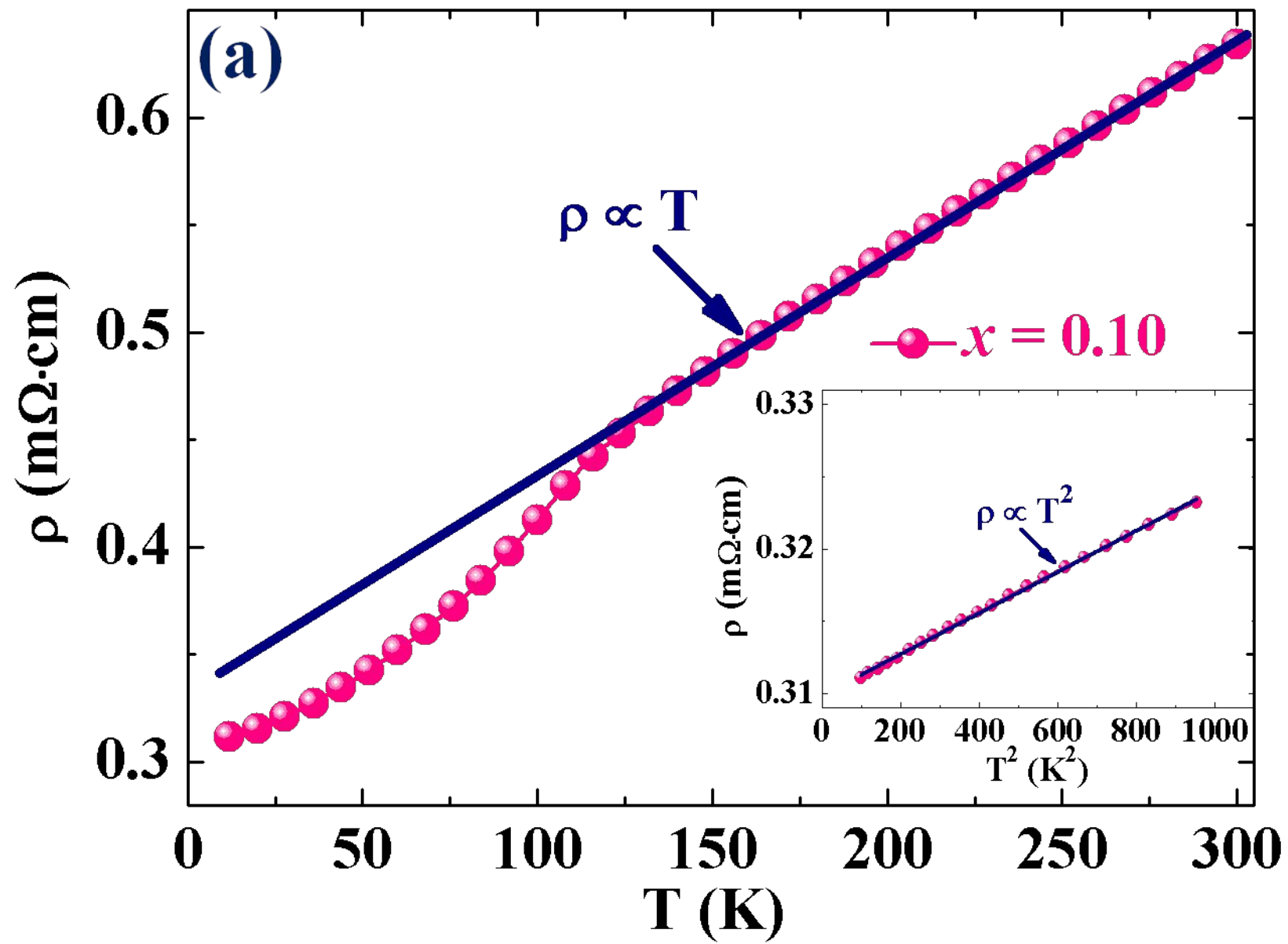
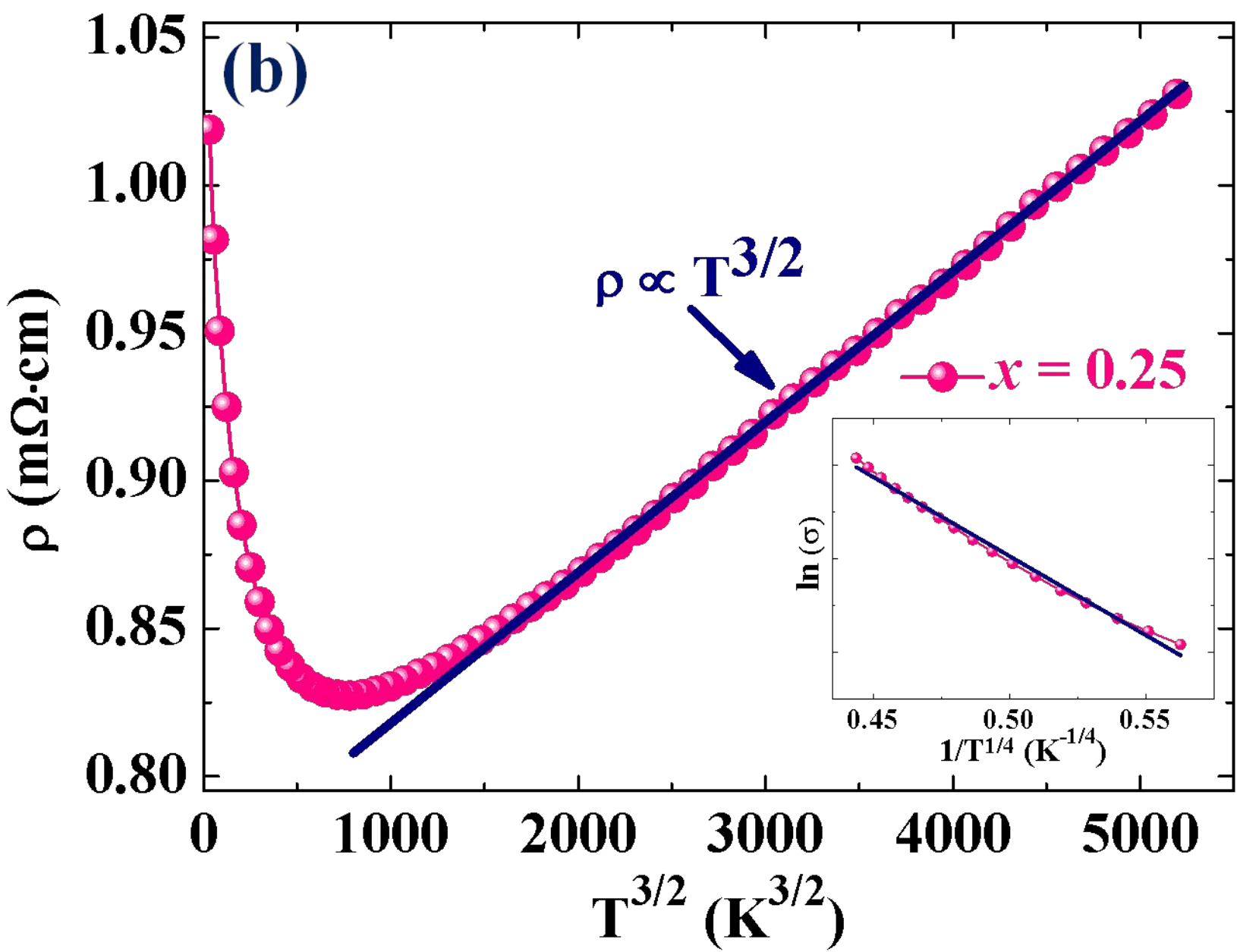

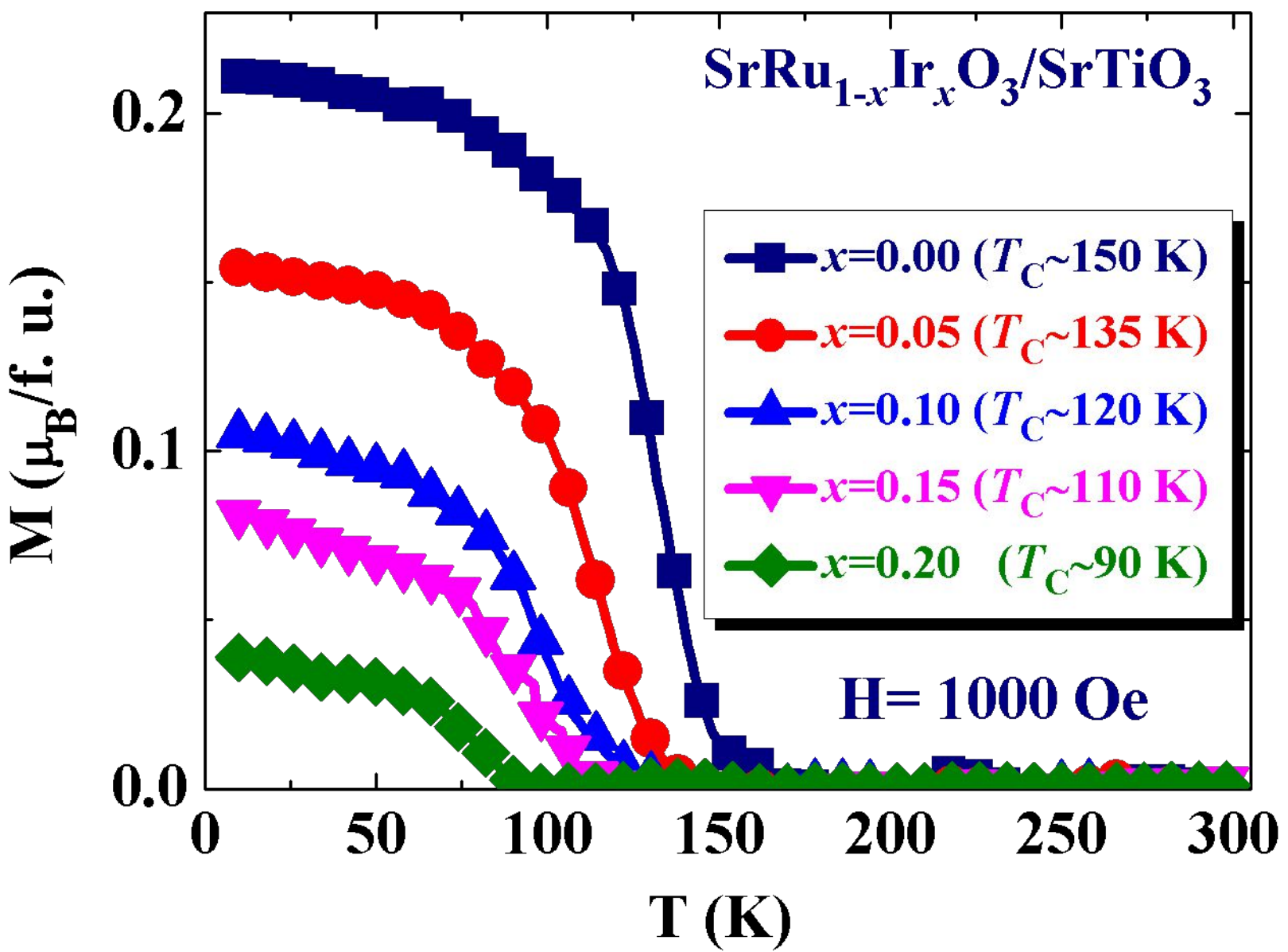

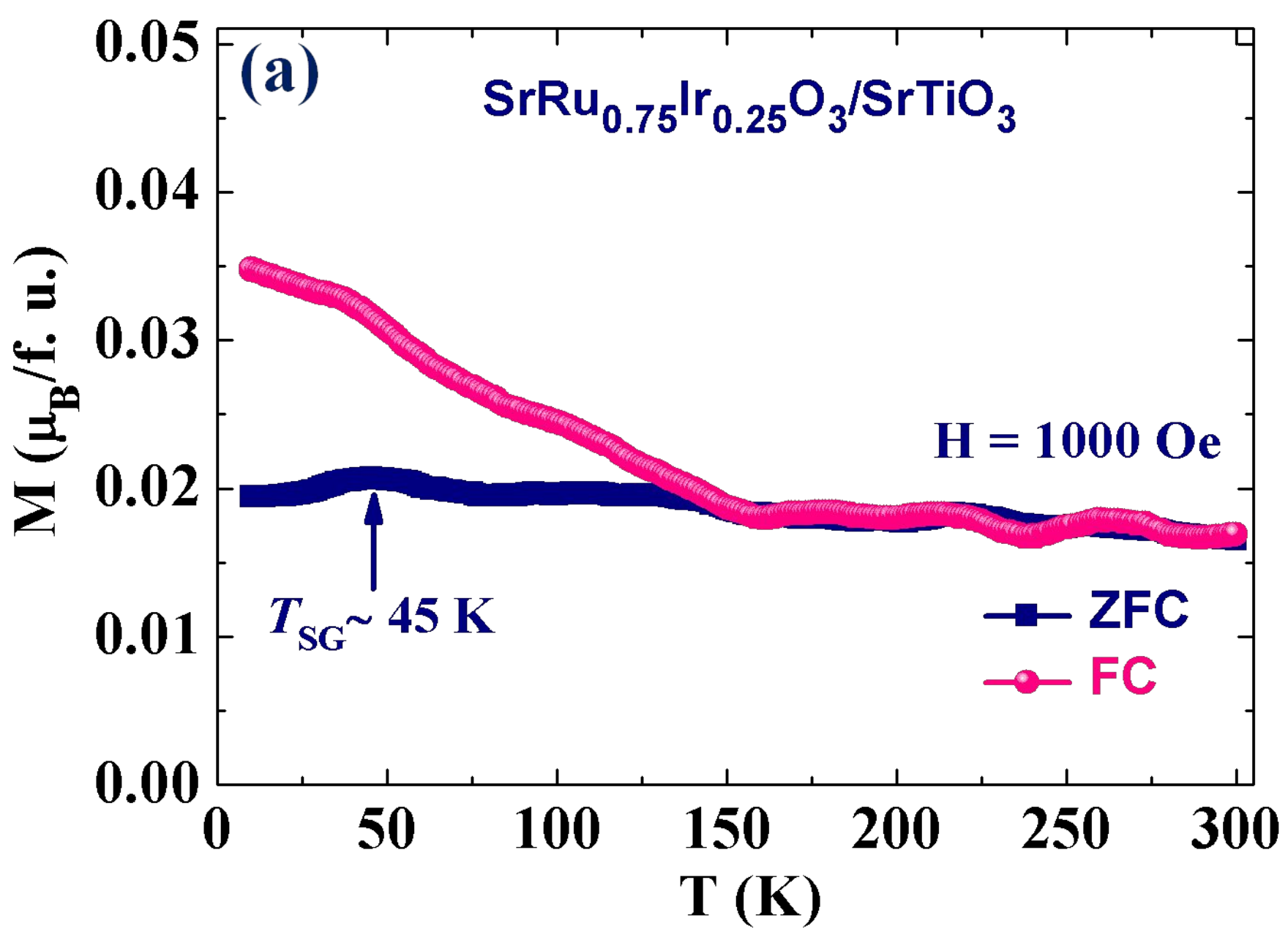
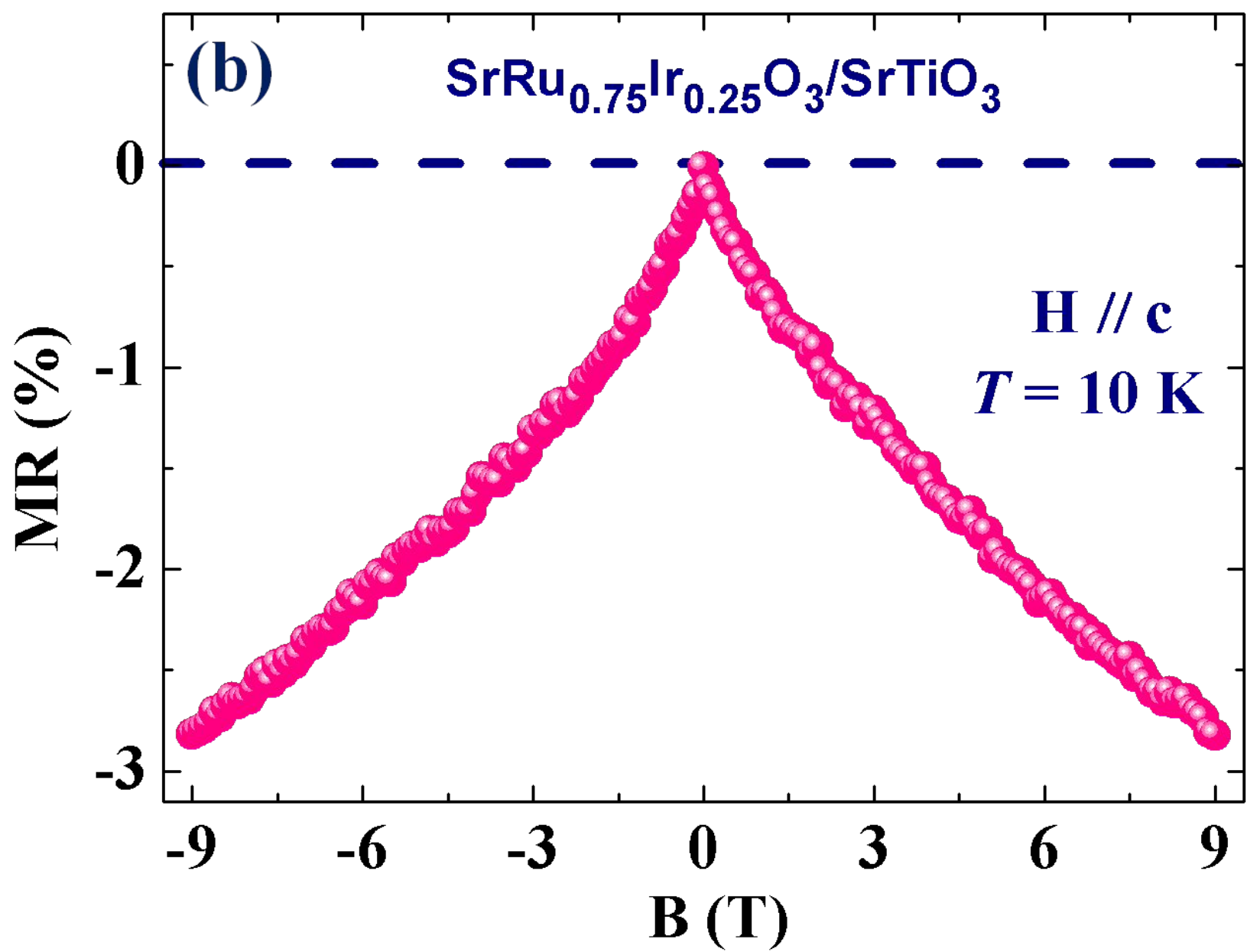